\documentclass[12pt,preprint]{aastex}
\usepackage{epstopdf}
\usepackage{graphicx}
\usepackage{spr-astr-addons}
\usepackage{url}
\usepackage{lscape}
\bibpunct{(}{)}{,}{n}{,}{,}

\RequirePackage{color}

\begin{document}

 \title{The Pleiades apex and its kinematical structure}

 \shorttitle{The Pleiades kinematics}
 \shortauthors{Elsanhoury et al.}

 \author{W. H. Elsanhoury}
 \affil{Astronomy Department, National Research Institute of Astronomy and Geophysics (NRIAG) 11421,
Helwan, Cairo, Egypt, (Affiliation ID: 60030681)}
  \and
 \affil{Physics Department, Faculty of Science, Northern Border University, Rafha Branch, Saudi Arabia}
  \and
 \author{E. S. Postnikova}
  \and
 \author{N. V. Chupina}
  \and
 \author{S. V. Vereshchagin}
 \affil{Institute of Astronomy, Russian Academy of Sciences, 48 Pyatnitskaya st., Moscow, Russia}
 \author{Devesh P. Sariya}
 \affil{Department of Physics and Institute of Astronomy, National Tsing Hua University, Hsin-Chu, Taiwan}
 \author{R. K. S. Yadav}
 \affil{Aryabhatta Research Institute of observational sciencES (ARIES), Manora Peak Nainital 263 002, India}
 \author{Ing-Guey Jiang}
 \affil{Department of Physics and Institute of Astronomy, National Tsing Hua University, Hsin-Chu, Taiwan}
\email{welsanhoury@gmail.com (W. H. Elsanhoury); es$_$p@list.ru (E. S. Postnikova), 
chupina@inasan.ru~(N.~V.~Chupina);
svvs@ya.ru~(S.~V.~Vereshchagin); deveshpath@gmail.com~(Devesh P. Sariya);
rkant@aries.res.in (R. K. S. Yadav); jiang@phys.nthu.edu.tw (Ing-Guey Jiang)}

\begin{abstract}
A study of cluster characteristics and internal kinematical structure 
of the middle-aged Pleiades open star cluster is presented. 
The individual star apexes and various cluster kinematical parameters including 
the velocity ellipsoid parameters are determined using both Hipparcos and Gaia data.  
Modern astrometric parameters were taken from the Gaia Data Release 1 (DR1) 
in combination with the Radial Velocity Experiment Fifth Data Release (DR5). 
The necessary set of parameters including parallaxes, 
proper motions and radial velocities are used 
for n=17 stars from Gaia DR1+RAVE DR5 and 
for n=19 stars from the Hipparcos catalog using SIMBAD data base. 
Single stars are used to improve accuracy by eliminating orbital movements. 
RAVE DR5 measurements were taken only for the 
stars with the radial velocity errors not exceeding $2$~km/s. 
For the Pleiades stars taken from Gaia, we found mean heliocentric distance as 
$136.8 \pm 6.4$~pc,  
and the apex position is calculated as: 
$A_{CP}=92^\circ.52\pm1^\circ.72$, 
$D_{CP}=-42^\circ.28\pm2^\circ.56$ by the convergent point method
and $A_0=95^\circ.59\pm2^\circ.30$ and $D_0=-50^\circ.90\pm2^\circ.04$ using AD-diagram method (n=17 in both cases). 
The results are compared with those obtained historically before the Gaia mission era.
\end{abstract}


\section{Introduction}
\label{Intro}

Famously called as ``the Seven Sisters'',
Pleiades open cluster (M45, Melotte 22 and NGC 1432) has attracted the observers since antiquity.
Over the past century, the stellar content of the cluster have been studied extensively. Pleiades
($\alpha$$_{2000}$ = 03$^{\rm h}$ 47$^{\rm m}$ 24$^{\rm s}$, $\delta$$_{2000}$ = $+24^\circ$ 07$'$ 00$''$)
is located in the constellation of Taurus and contains about 
two thousand cluster members.
Bouy et al. (2015) applied a probabilistic method based on multivariate data analysis 
(see, Sarro et al., 2014) to select high probability members of the Pleiades on the
wide field data taken in multi epoch (Bouy et al., 2013) and Tycho-2 (H{\o}g et al. 2000) catalogs.
They identified a total of 2107 high probability member stars in the Pleiades region
and produced the most complete member list.

The distance of Pleiades has been an interesting puzzle for the astronomical community.
To date, there are several estimates of the heliocentric distance of Pleiades 
obtained by different methods. 
Using the moving cluster method, Galli et al. (2017) gave the distance value as $134.4^{+2.9}_{-2.8}$~pc. 
Melis et al. (2014) determined the distance of $136.2 \pm 1.2$~pc 
to the Pleiades based on Very Long Baseline Interferometry (VLBI).
Brown et al. (2016, Gaia collaboration) using the data from Gaia Data Release~1 (Gaia DR1) obtained 
the cluster's mean parallax value as $7.45\pm0.3$~mas corresponding to a distance of 
about $134\pm6$~pc.  
The method of double stars orbital modeling 
led to cluster's distance value of $138.0 \pm 1.5$~ pc (Groenewegen et al. 2007). 
Recently, the twin stars method yielded the distance of $134.8 \pm 1.7$~pc to the Pleiades open cluster ( M\"adler et al. 2016). 
As can be noticed, majority of the distance estimates provide similar results, close to $\sim$135~pc. 
However, van Leeuwen (2009), 
based on the re-calibrated parallaxes of 53 stars of the Hipparcos data (ESA, 1997), 
derived a distance $120.2 \pm 1.9$~pc,
which differs from the above estimates. 
A possible reason for this underestimation can be the technique for averaging parallaxes. 
There was a long standing problem regarding the distance of Pleiades. 
According to Hipparcos data, the Pleiades distance is 118 pc, 
while most of the other determinations were in the range of 130 to 140 pc,
(Gaia Collaboration, van Leeuwen 2017).
The difference in distance measurement secured using Hipparcos data 
may be due to the random errors (Francis \& Anderson 2012) and other undetected causes. 

The purpose of this article is to study the kinematical parameters of the Pleiades cluster. 
In this analysis, we have derived apex coordinates of the cluster using the 
cluster members selected on the basis of proper motions and radial velocity data. 
We are using two methods to determine convergent point of Pleiades: 
the Convergent Point (CP) method based on proper motions 
and the AD-chart method based on spatial velocity vectors. 
Depending on the accuracy of available astrometric measurements, 
the methods used in this work can provide precise apex position.
We have also derived the cluster spatial velocity, 
velocity ellipsoid parameters etc. based on different astrometric data
taken from Gaia DR1 and Hipparcos catalogues.

This paper is organized as follows. In Sect. 2, we describe the sample of Pleiades stars and
the dataset (proper motions, radial velocities and parallaxes) used in this analysis.
In Sect. 3, we describe derivation of apex coordinate and different formulae. 
Discussions on different parameters are presented in Sect. 4. 
Finally, we summarize the various results in Table~3.

\section{Sample of stars}
\label{OBS}

In this analysis, the main results are based on the recent data
contained in the Tycho-Gaia astrometric solution (TGAS) of Gaia DR1 catalog
(Brown et al. 2016). Recently, van Leeuwen et al. (Gaia collaboration, 2017) 
used the TGAS data
for some open clusters including the Pleiades.
We used the Pleiades stars' list from van Leeuwen (Gaia collaboration, 2017)
and combined this data with radial velocities from the 
fifth Data Release of the Radial Velocity Experiment (RAVE DR5, Kunder et al. 2017).
In order to understand the change from the old data based results, 
we also carried out the calculations with data from 
the Hipparcos catalog (HIP, ESA 1997)
and the New Hipparcos astrometric catalog (HIP New), (van Leeuwen 2007).
Since our goal is to obtain the most accurate cluster parameters based on astrometry, 
in all the cases, we used data only for single stars, 
eliminating the double and multiple systems 
whose parameters can be distorted by orbital motions. 
This also applies to the radial velocities data. 
In the case of both versions of the Hipparcos catalog, 
we took radial velocities 
from the SIMBAD\footnote{http://simbad.u-strasbg.fr/simbad/}
(Wenger et al. 2000) database. 
The individual sources of radial velocities in SIMBAD are  
Wilson (1953); Gontcharov (2006); Kharchenko et al. (2007); Mermilliod et al. (2009); 
Kordopatis et al. (2013) and Kunder et al. (2017). 

In van Leeuwen (2009), 57 stars were found in the Pleiades region, 
from which he dropped four stars which were double stars with high orbital motion velocity. 
This data was not published with the paper and Dr. van Leeuwen kindly provided us the data
on our request.
We used this list as a base for the star parameters search and calculations.  
From the rest of the 53 stars, we could find the $V_r$ values for 48 stars in the SIMBAD. 
As already mentioned, for the HIP and HIP New data, we had at our disposal, 
two samples of parallaxes and proper motions data with their errors. 
These two samples are identical in stars and their radial velocities. 
Table~1 lists the stars with two sets of astrometric parameters for the 
two mentioned catalogues.

Table~1 shows our sample of 48 stars with known radial velocities.
The columns in Table 1 are: star number from the HIP catalog,
the parallaxes and their errors (in mas), proper motions and their errors (in mas/yr) 
from HIP and HIP New, 
the radial velocity $V_r$ and its error $\sigma_{V_r}$ (in km/s) found in SIMBAD. 
The flags d1 and d2 when denoted by the ``+'' mark, represent double stars. 
Here, d1 comes from the HIP parameters.
When the Catalog of Components of Double and Multiple stars (CCDM) identifier number in column
H55 and/or H59 is non-empty, the flag of that star is `double'.
Also, if the identifier H61 shows a flag `S', that means the star is suspected to be non-single.
The flag d2 is taken from the Washington Double Star 
(WDS)\footnote{http://ad.usno.navy.mil/wds/} catalog.
HIP data is used to illustrate the errors in the AD-diagram.
HIP contains correlation coefficients necessary for constructing the error ellipses.

We checked the double star flag (d1,d2) as marked 
in the Table~1 which left us with 33 stars.
The star number HIP~17401 was excluded as its $V_r$ value ($-46.900$ 
km/s) is outside the three sigma limit. 
Finally, we applied a cut-off of 2~km/s on the radial velocity error,
which reduced our sample to 19 stars.
To calculate the equatorial coordinate of the position of the average apex, 
we used these 19 single stars.

We used the sample of TGAS stars presented by van Leeuwen et al. (2017, Gaia collaboration). 
We provide a list of 35 stars in Table~2, for which we 
were able to find the radial velocities in RAVE DR5  (Kunder et al. 2017). 
If there were more than one measurements of radial velocity,  
we chose the value with the smallest relative error.

Table~2 columns indicate the star number from the TGAS catalog, the RA and DEC in ICRS,
epoch $=$ J2015.0, the parallaxes and their errors (mas), proper motions and their errors (mas/yr)
from TGAS, the radial velocity and its errors (in km/s) from RAVE DR5,
and the double star flags d1 and d2.
Among the single stars in Table~2, there are two stars with 
radial velocities significantly exceeding the average value for the Pleiades cluster.
These stars are Gaia 5754128236100096 ($V_r=44.166$~km/s) 
and Gaia 67618281484716544 ($V_r=109.263$~km/s).
For 13 stars, the radial velocity error exceeds 2.0~km/s. 
These 15 stars and 3 double stars are excluded from our calculations. 
Conclusively, we used 17 single stars from Table~2.

\section{Kinematical analysis}
\label{kin}
\subsection{Apex of the moving cluster}

We study the kinematics of Pleiades using HIP, HIP New and TGAS data. 
Using these data sets, we have calculated
the apex (vertex or convergent point) position of the cluster and 
velocity ellipsoid parameters.

By virtue, the stars associated with an open cluster share the similar parameters like age, distance, kinematics and
chemical composition. Our aim is to determine the point at which the motion of stars in the cluster will converge,
i.e. vertex of Pleiades. For this, following two methods can be used:\\

\begin{enumerate}
\item The CP-method
\item The AD-chart method
\end{enumerate}

Results using different data inputs and based on the two methods applied 
independently are given below.

\subsubsection{The convergent point method}

The method of the convergent point is a classical method for identifying 
the cluster membership of stars with the help of proper motions vectors. 
It has been used for almost a century. 
In the consecutive years, the method has been explained and refined by several authors,
e.g., Jones (1971); de Bruijne (1999); Galli et al. (2012) etc. 
This method allows to select stars by evaluation of parallelism of the components of proper motions. 

For a group of $N$ cluster member stars with coordinates ($\alpha,\delta$), 
located at a distance $r_i$ (pc), proper motions in RA and DEC, $\mu_{\alpha}cos\delta$ and
$\mu_{\delta}$ (mas/yr) and radial velocity, 
$V_r$ (km/s) are used in this analysis.
Considering the above parameters, we can estimate the velocity components $(V_x, V_y, V_z)$ 
along $x,y$ and $z$ axes in the coordinate system centered at the Sun (see, Elsanhoury et al. 2015).
In this coordinate system, the plane of the equator is taken as the main coordinate plane. 
The main axis of the $OX$
reference is directed from the origin $O$ to the vernal equinox $T$.
The $OY$ axis is at an angle of 90 degree to the $OT$ axis. 
The OZ axis is perpendicular to the other axes and points to the North pole.

According to the well-known formulae given by Smart (1968):
\begin{center}
$$
\arraycolsep=0.1pt
\begin{array}{lll}
V_x &=& -4.74r_i\mu_\alpha\cos\delta\sin\alpha-4.74r_i\mu_\delta\sin\delta\cos\alpha\\
&&+V_r\cos\delta\cos\alpha, \\
V_y &=& +4.74r_i\mu_\alpha\cos\delta\cos\alpha-4.74r_i\mu_\delta\sin\delta\sin\alpha\\
&&+V_r\cos\delta\sin\alpha, \\
V_z &= &+4.74r_i\mu_\delta\cos\delta+V_r\sin\delta. \\
\end{array}
$$
\end{center}

From the above equations and letting
\begin{center}
$\xi  = \frac{V_x}{V_z},$ \\
$\eta = \frac{V_y}{V_z},$ \\
\end{center}

we get
\begin{center}
$a_i \xi+b_i \eta  = c_i. $ \\
\end{center}

where the coefficients
\begin{center}
$a_i = \mu_\alpha^{(i)}\sin\delta_i\cos\alpha_i\cos\delta_i-\mu_\delta^{(i)}\sin\alpha_i,$ \\
$b_i = \mu_\alpha^{(i)}\sin\delta_i\sin\alpha_i\cos\delta_i+\mu_\delta^{(i)}\cos\alpha_i,$ \\
$c_i = \mu_\alpha^{(i)}\cos^2\delta_i.$ \\
\end{center}

the index $i$ varies from 1 to $N$, which is the number of cluster members. So, we have
the following equations for the cluster's vertex ($A_{CP}, D_{CP}$):
\begin{center}
$\tan A_{CP}  = \frac{\eta}{\xi}, $ \\
$\tan D_{CP}  = \frac{1}{\sqrt{\eta^2+\xi^2}}. $ \\
\end{center}

The velocity dispersion (i.e. internal motions within the cluster) 
was not taken into account in this study the way it was used in other methods 
like de Bruijne (1999); Galli et al. (2012). 
This would mean that the so-derived uncertainties 
in the apex positions are underestimated.
It needs to resolve the selection of stars 
with imperfectly null vectors of proper motions, i.e.
the components perpendicular to the direction of apex. 
For the present task, this was not relevant.

In Table~3, we present the coordinates of the apex position  $A_{CP}$ and $D_{CP}$. 
In this task, we used 19 stars from 
Table~1, HIP New (HIP numbers marked by asterisks)
and 17 stars from Table~2 
(TGAS IDs marked by asterisks).

\subsubsection{The AD-chart method}

The number of stars with available measurements of the proper motions is usually much higher than the 
number of available stars with known radial velocity values.  
However, for the complete picture of the stars' space motions, both proper motions and space veloctities are required. 
By including the radial velocities to identify stellar groupings that have a common movement in space, 
a stellar apex method was developed, known as the AD-diagram method.

This method has been used and discussed for the kinematics 
of Ursa Major kinematic stream (Chupina et al. 2001, 2006). 
This method is also used to study the Hyades (Vereshchagin et al. 2008), 
Orion Sword region (Vereshchagin \& Chupina 2010), 
Praesepe (Vereshchagin \& Chupina 2013), M 67 (Vereshchagin et al. 2014), 
Castor moving group (Vereshchagin \& Chupina 2015), 
and recently, NGC 188 (Elsanhoury et al. 2016).

The AD-diagram method uses the notion of an ``individual stellar apex". 
This term was introduced by analogy with the apex of the Sun or the cluster's apex -- 
the point on the celestial sphere, towards which the object moves, in this case, a star. 
Individual apexes can be obtained, if for the vector of the space velocity of the star, 
the beginning is placed on the point of observation and it is extended towards 
the intersection with the surface of the celestial sphere. 
The intersection point is the desired apex. 
AD-diagram shows the positions of individual apexes in equatorial coordinates. 
The proximity of the locations of the points on the AD-diagram indicates the parallelism of 
the corresponding spatial velocity vectors. 
By condensing points on the diagram, 
we can distinguish groups of stars having a common motion in space. 
This method is convenient for its simplicity and clarity. 
Unlike the $(UVW)$ diagrams, where the speed ellipsoid needs to be considered, 
the stellar apex method allocates co-directional vectors on the 2D plane, 
with no restrictions imposed on the modulus of the velocity vector.

A formal description of the method, diagramming technique and
formulae to determine the error ellipses can be found in Chupina
et al. (2001). This method requires knowledge of radial velocity
and parallax. 
Note that the error ellipse can be determined 
for HIP data, but not for the HIP New stars as they do not have the required
correlation between astrometric parameters.
Gaia DR1 (TGAS) also provides the full covariance matrix 
for the stars with measured parallaxes and proper motions and allows to construct 
the error ellipses on the AD-diagram. 
We have constructed the error ellipses for both HIP and Gaia data, as they are 
shown in the panels (b) and (c) of Fig.~1.
More about this is discussed in Section 3.1.5.
The AD-diagram could be used to study the kinematical structure 
of the cluster and to find out cluster's inner kinematical substructures.

\subsubsection{The ($A_0,D_0$) formulae}

Recently, Vereshchagin et al. (2014) and Elsanhoury et al. (2016) used the
method explained by Chupina et al. (2001, 2006) for apex determination. In this method,
equatorial coordinates of the convergent point are calculated as:
\begin{center}
$A_0 = \tan^{- 1}(\frac{\overline{V_y}}{\overline{V_x}}), $ \\
$D_0 = \tan^{-1}(\frac{\overline{V_z}}{\sqrt{\overline{V^2_x}+\overline{V^2_y}}}).
$
\end{center}

\subsubsection{The ($AD$) error ellipses}

The technique to determine various coefficients and formulae for
($AD$) error ellipses are explained in Brown et al. (1997); Chupina et al. (2001). 
The formulae are given as follows:\\

$$Cov{{\rm A}\choose{\rm D}}={\rm JCJ}^{\rm T}$$

where C is the proper motions and parallax covariation matrix

\begin{center}
$$
\arraycolsep=0pt
\begin{array}{cccc}
{\rm C} &=& \left(
\begin{array}{cccc}
\sigma^2_{\mu_\alpha}&
\sigma_{\mu_\alpha}\sigma_{\mu_\delta}\rho_{\mu_\alpha\mu_\delta}&
\sigma_{\mu_\alpha}\sigma_{\pi}\rho_{\mu_\alpha\pi}&0\\
\sigma_{\mu_\alpha}\sigma_{\mu_\delta}\rho_{\mu_\alpha\mu_\delta}&
\sigma^2_{\mu_\delta}&
\sigma_{\mu_\delta}\sigma_{\pi}\rho_{\mu_\delta\pi}&0\\
\sigma_{\mu_\alpha}\sigma_{\pi}\rho_{\mu_\alpha\pi}&
\sigma_{\mu_\delta}\sigma_{\pi}\rho_{\mu_\delta\pi}&
\sigma^2_{\pi}&0\\
0&0&0&\sigma^2_{{\rm V}_{\rm r}}\\
\end{array}
\right),
\end{array}
$$
\end{center}

and J the Jacobian to transform matrix
$\left[
\begin{array}{c}
\mu_\alpha\\\mu_\delta \\ \pi \\ V_r
\end{array}
\right] $
to the matrix
$\left[
\begin{array}{c}
 A \\ D
\end{array}
\right] $
and is as following:

\begin{center}
$$
\begin{array}{ccc}
{\rm J}&=&
\left(
\begin{array}{cccc}
\frac{\displaystyle\partial{\rm A}}{\displaystyle\partial\mu_\alpha}&
\frac{\displaystyle\partial{\rm A}}{\displaystyle\partial\mu_\delta}&
\frac{\displaystyle\partial{\rm A}}{\displaystyle\partial\pi}&
\frac{\displaystyle\partial{\rm A}}{\displaystyle\partial{\rm V}_{\rm r}}\\
\noalign{\smallskip}
\frac{\displaystyle\partial{\rm D}}{\displaystyle\partial\mu_\alpha}&
\frac{\displaystyle\partial{\rm D}}{\displaystyle\partial\mu_\delta}&
\frac{\displaystyle\partial{\rm D}}{\displaystyle\partial\pi}&
\frac{\displaystyle\partial{\rm D}}{\displaystyle\partial{\rm V}_{\rm r}}\\
\end{array}
\right).
\end{array}
$$
\end{center}

The confidence region around the $(A,D)$ is given by the formula

\begin{center}
$$ C = \left(\Delta{\rm A,}\, \Delta{\rm D}\right) \left[Cov{{\rm
A}\choose{\rm D}}\right]^{-1} {\Delta{\rm A}\choose\Delta{\rm D}}
. $$
\end{center}

We use covariance matrix C for the calculations of errors ellipses
shown in Fig. 1. The coefficient C is equal to 11.83 for number of
degree of freedom being two (see, Brown et al. 1997).

\subsubsection{The AD-diagrams for different catalogs}

Figure~1 represents the AD-diagrams made with different astrometric data. 
We have separately plotted AD-diagrams taking 
only the single stars (left panels) and the 
double and multiple stars (right planes).
This exhibits the difference in results when the astrometry data 
is not burdened by orbital movements in binary and multiple systems.
Ultimately, only the single stars were used in all our calculations.
In the top panel of Fig.~1 marked as (a), results based on the 
parallaxes and proper motions taken from HIP New data (Table~1) 
are shown. The middle panel (b) shows the results obtained from the HIP data (Table~1). 
The bottom panel (c)  contain the AD-diagram calculated using TGAS sample (Table~2). 
The data in Table~1 (based on HIP data) and Table~2 (on TGAS) 
made it possible to construct error ellipses.
The choice of the order of panels is based on the ability to calculate the error ellipses.
This is why HIP data is plotted in the panel (b), close to TGAS data (panel c)
as they both have the error ellipses in their AD-diagrams. 
The information about the number of stars used in different panels is presented in Table~4.
The filled dots in red color (Fig.1) represent the stars whose
radial velocity errors do not exceed 2~km/s. 
As mentioned earlier, only these stars were used
to calculate the coordinates of the cluster apex.

It was deemed necessary to show AD-diagram according to the Hipparcos data in Fig.~1, 
even though they are less accurate than HIP New in order to see the error ellipses. 
Please recall that the correlation coefficients necessary for building error ellipses 
exist in the HIP, but not in the HIP New. 
The radial velocities for the panels (a) and (b) are identical, as found in SIMBAD. 
The bottom panel of the Fig.1 shows the main result which represents 
the diagram plotted using the Gaia astrometric parameters 
and radial velocities from RAVE DR5. 

The results of apex position calculation using the data listed in Table~1, HIP New  (used n=19)
and Table~2 (used n=17) are presented in Table~3. 
Table~3 contains the coordinates of the cluster apex position 
with their mean square errors ($\sigma_{A_0}$, $\sigma_{D_0}$) 
received by us and other authors for the comparison.
The formulae for ($A_0$, $D_0$) and the error ellipses for star apexes
are discussed in the Sections 3.1.3 and 3.1.4. 

The apex values calculated using different data sources 
do not differ significantly both in position and r.m.s. scatter. 
Also, these results show that 
the recently released Gaia data is more accurate than Hipparcos data. 
Points on the bottom panel in the Fig. 1 are distributed 
in more compact manner as compared to the top two panels
but not in all the directions, i.e., along the inclined line, the scatter remains large. 
Points with errors in radial velocities exceeding 2 km/s are located in the trails of the diagrams,
which are the farthest from the average position of the apex.

Figure~2 represents a comparison of parallaxes taken from HIP New and TGAS.
We see that there is significant difference between the individual parallaxes   
determined using the HIP New and TGAS.
The scatter of points and individual errors for the HIP New data
are several times higher than the corresponding values for TGAS data. 
Also, the values for the average parallax differ.
The most recent Gaia data (Brown et al. 2016) 
confirmed the non-Hipparcos results of the Pleiades distance interval. 
The difference in parallax  (Fig.~2) led to the differences in apex coordinates, 
but it is within $1^\circ$ as seen in Table~3. 
We used the TGAS data from Table~2 to determine 
the average parallax of individual stars. 
For n=17 single stars in the Pleiades, the average parallax value was converted into the
mean heliocentric distance of $136.8 \pm 6.4$~pc.

\subsection{The cluster spatial velocity (relatively to the Sun)}

The velocity of the cluster can be calculated by the following formula:
$$V = \frac{\sum_{i=1}^{N} V_r^{(i)}\cos\lambda_i}{\sum_{i=1}^{N}\cos^2\lambda_i}, $$
where $\lambda$ is the angular distance from the star to the vertex:

\begin{center}
$\lambda_i = \cos^{- 1}[\sin\delta_i\sin D_0+\cos\delta_i\cos D_0\cos(A_0-\alpha_i)]. $ \\
\end{center}

The results of calculations of the positions of the cluster apex by different methods, 
as well as the obtained kinematical parameters and the parameters of velocity ellipsoid are given in Table~3. 
Our values of the components of the spatial velocity 
of the cluster relative to the Sun $(V_x, V_y, V_z)$ are shown in Table~3. 
In order to compute components of space velocity in the Galactic space coordinates $(U, V, W)$, 
we used the tranformations given by Murray (1989).

\subsection{The velocity ellipsoid parameters}

To compute the parameters of the ellipsoidal velocities for the
members of Pleiades, we used a computational algorithm (see,
Elsanhoury et al. 2015). The coordinates, $Q_i$,  of the point $i$
with respect to an arbitrary axis $\xi$ centered on the stellar
distributions center are determined. Then, the algorithm is used
to calculate generalized form of the mean square deviation
$\sigma^2$, the direction cosines vector $B$, a symmetric matrix
with elements $\mu_{ij}$ and eigenvalues $\lambda_{1,2,3}$.

Now, we establish analytical expressions of some parameters for
the correlation studies in terms of the matrix elements $\mu_{ij}$
of the eigenvalue problem for the Velocity Ellipsoid Parameters
(VEPs).
\\

\noindent $\bullet$ The {$\sigma _{i} ;{\rm \; }i=1,2,3$}
parameters are defined as
\begin{equation}
\sigma _{i} = \sqrt{\lambda _{i} }.
\end{equation}

\noindent $\bullet$ The $l_{i} ,{\rm \; }m_{i} {\rm \; and\;
}n_{i} $ are the direction cosines for eigenvalue problem. Then we
have the following expressions for {$l_{i} ,{\rm \; }m_{i} {\rm \;
and\; }n_{i} $} as
\begin{equation}
l_i =\frac{\mu _{22} \mu _{33} -\sigma _i^2\left(\mu _{22} +\mu _{33} -\sigma _i^2\right)-\mu _{23}^2}{D_i},\; i=1,2,3
\end{equation}
\begin{equation}
m_i =\frac{\mu _{23} \mu _{13} -\mu _{12} \mu _{33} +\sigma _{i}^{2} \mu _{12}}{D_i}, \; i=1,2,3
\end{equation}
\begin{equation}
n_i =\frac{\mu _{12} \mu _{23} -\mu _{13} \mu _{22} +\sigma _{i}^{2} \mu _{13}}{D_i}, \; i=1,2,3
\end{equation}
where
$$
\small
\arraycolsep=0.1pt
\begin{array}{lll}
D_i^2&=&\left(\mu _{22} \mu _{33} -\mu _{23}^{2} \right)^{2} +
\left(\mu _{23} \mu _{13} -\mu _{12} \mu _{33} \right)^{2} +\\
&&\left(\mu _{12} \mu _{23} -\mu _{13} \mu _{22} \right)^{2} +
2[\left(\mu _{22} +\mu _{33} \right) \left(\mu _{23}^{2} -\mu _{22} \mu _{33} \right)+\\
&&\mu _{12} \left(\mu _{23} \mu _{13} -\mu _{12} \mu _{33} \right)+
\mu _{13} \left(\mu _{12} \mu _{23} -\mu _{13} \mu _{22} \right)]\sigma _i^2+ \\
&&\left(\mu _{33}^{2} +4\mu _{22} \mu _{33} +\mu _{22}^{2} -2\mu _{23}^{2} +\mu _{12}^{2} +\mu _{13}^{2} \right)\sigma _i^4-\\
&&2\left(\mu _{22} +\mu _{33} \right)\sigma _i^6 +\sigma _i^8.\\
\end{array}
$$
\label{sec:math}

The results of velocity ellipsoid parameters calculations are given in the Table~3. 
In total, n=19 single stars from Table~1, with $\sigma_{V_r} \le 2.0$~km/s were used for HIP New 
and n=17 stars for the TGAS data (Table~2) calculations.

\section{Discussion and Conclusions}
\label{con}

Pleiades remains of great importance for understanding the stellar kinematics.
Some parameters similar to this work were determined 
by Chen et al. (1999); Makarov \& Robichon (2001); Galli et al. (2017).
Recent data with accurate measurements of astrometric parameters 
and radial velocities motivated us to carry out the present work.

The apex coordinate and various kinematical structure parameters of the 
Pleiades open cluster have been studied. 
A computational routine using the ``Mathematica'' software has 
been developed to compute the kinematical parameters of this cluster.
We calculated the coordinate positions by two independent methods: AD-diagram and CP-method.
Using HIP New data, we obtained 
($A_{CP},D_{CP})=(95^\circ.73\pm3^\circ.56, -50^\circ.44\pm8^\circ.84)$ by the CP method
and using the AD-diagram $(A_0, D_0) = (93^\circ.06\pm5^\circ.95, -48^\circ.42\pm4^\circ.02)$ was calculated.
Using the TGAS and RAVE DR5 data as the input, we determined the apex values as: 
($A_{CP},D_{CP})=(92^\circ.52\pm1^\circ.72, -42^\circ.28\pm2^\circ.56)$ by the CP-method 
and ($A_0, D_0) = (95^\circ.59\pm2^\circ.30, -50^\circ.90\pm2^\circ.04)$ using the AD-diagram. 
Different kinematical parameters are derived considering HIP New data and
newly acquired Gaia astrometric data and the results are given in Table~3. 
Only single stars with $\sigma_{V_r}\le2.0$~ km/s are used (red color points in Fig.~1).
Results from both the AD-diagram and CP methods are almost similar. 
As expected, TGAS shows better accuracy 
of input data in determining the positions of the apexes and other parameters (Table~3). 
Our results of the determination of apex coordinates are in good agreement 
with the results of other authors (Makarov \& Robichon 2001; Galli et al. 2017), 
which are also mentioned in Table~3. 
There is a discrepancy with the results obtained by Montes et al. (2001), 
which could be because of the fact that they used the ESA HIP (1997) data.

The present cluster distance determination using TGAS+RAVE DR5 
is in good agreement with previous results, 
but the distance value differs when, as was shown in Sec. 3.1.5, 
we used HIP New data given in Table~1. 
However, TGAS high accuracy could not be fully utilized. 
The reason is that the accuracy of the radial velocities is not yet comparable with Gaia astrometric accuracy, 
even for the RAVE DR5 catalog used here. 
For the same reason, the AD-chart (for Gaia data) presented in Fig.~1 has an elongated form. 
It is stretched both according to the positions of the individual points, 
and over the areas of error ellipses. 
It is obvious that the results from the Gaia data are more accurate and
we will consider them as the main result of this work. 
Results from both the AD-diagram and CP methods are almost similar. 
We have shown that the kinematical parameters derived using 
Gaia DR1 (TGAS) data are more reliable than other data sets. 
But, this data set with a combination of Gaia DR1 and RAVE DR5 does not contain too many stars (35 stars, Table~2). 
Thus, the Gaia DR1 data did not provide an advantage in terms of the number of stars 
found in the list of Pleiadian membership (Table~2). 
This fact made it impossible to study the fine structure of the AD-diagram 
and to identify possible kinematical groups within the cluster as we did for M~67 in Vereshchagin et. al. (2014).

In this paper, we determined various kinematical parameters of Pleiades
using Gaia DR1 and Hipparcos data.
As the main result, we derived the position of the apex for the cluster
considering two different methods, the CP point and the AD method. 
We compared the AD-diagrams constructed from various data sources, 
i.e. HIP, HIP New and TGAS. 
The current sample of stars is not sufficient to  
enable the analysis of the structure of AD-diagram. 
The unprecedented high accuracy of the next data release of GAIA data (Gaia DR2) 
is expected to define the distribution of apexes in the AD diagram in a better way. 
This will also refine the details of the kinematical 
structure of the Pleiades and increase the number of Pleiades stars for consideration
(Katz \& Brown 2017).

\section{Acknowledgments}

We are thankful to the referee of this paper for many useful comments 
including the data source which highly improved the level of this paper.
We are thankful to Dr. Floor van Leeuwen for providing HIP New Pleiades members star list. 
This work has made use of data from the European Space Agency
(ESA) mission Gaia (http://www.cosmos.esa.int/gaia), processed by
the Gaia Data Processing and Analysis Consortium (DPAC,
http://www.cosmos.esa.int/web/gaia/dpac/ consortium). Funding for
the DPAC has been provided by national institutions, in particular
the institutions participating in the Gaia Multilateral Agreement.
This research has made use of the SIMBAD database, operated at
CDS, Strasbourg, France. This research has made use of the WEBDA
database, operated at the Department of Theoretical Physics and
Astrophysics of the Masaryk University. This research also made use of the RAVE data.
Funding for RAVE has been provided by: the Australian Astronomical Observatory; 
the Leibniz-Institut fuer Astrophysik Potsdam (AIP); the Australian National University; 
the Australian Research Council; the French National Research Agency; 
the German Research Foundation (SPP 1177 and SFB 881); the European Research Council 
(ERC-StG 240271 Galactica); the Istituto Nazionale di Astrofisica at Padova; 
The Johns Hopkins University; the National Science Foundation of the USA (AST-0908326); 
the W. M. Keck foundation; the Macquarie University; the Netherlands Research School for Astronomy; 
the Natural Sciences and Engineering Research Council of Canada; the Slovenian Research Agency; 
the Swiss National Science Foundation; the Science \& Technology Facilities Council of the UK; 
Opticon; Strasbourg Observatory; and the Universities of Groningen, Heidelberg and Sydney.
N. V. Chupina, S. V. Vereshchagin and E. S. Postnikova are partly
supported by the Russian Foundation for Basic Research (RFBR,
grant number is 16-52-12027). 
Devesh P. Sariya and Ing-Guey Jiang acknowledge the grant from Ministry of
Science and Technology (MOST), Taiwan. The grant numbers are MOST
103-2112-M-007-020-MY3, MOST 104-2811-M-007-024, and MOST
105-2811-M-007-038.


\clearpage
\begin{table*}
\footnotesize
\tabcolsep=0.07cm
\caption{Astrometric parameters for 48 HIP Pleiades cluster stars.
The stars whose HIP IDs are marked with asterisks are used for calculating the apex.
}
\vspace{0.5cm}
\centering
\begin{tabular}{lrcccccrcccccrcccr}
\hline\hline
&\multicolumn{6}{|c|}{HIP}&\multicolumn{6}{c|}{HIP New}&&&&\\
\multicolumn{1}{c}{HIP}  &  \multicolumn{1}{|c}{$\pi$}& $\sigma_\pi$ & $\mu_{\alpha}cos(\delta)$ & $\sigma_{\mu_{\alpha}}$& $\mu_{\delta}$& $\sigma_{\mu_{\delta}}$& \multicolumn{1}{|c}{$\pi$}& $\sigma_\pi$ & $\mu_{\alpha}cos(\delta)$ & $\sigma_{\mu_{\alpha}}$& $\mu_{\delta}$& $\sigma_{\mu_{\delta}}$& \multicolumn{1}{|c}{$V_r$} & $\sigma_{V_r}$ &d1&d2&\\
\multicolumn{1}{c}{ID}   & \multicolumn{2}{|c}{(mas)}& \multicolumn{2}{c}{(mas/yr)}  & \multicolumn{2}{c|}{(mas/yr)}   & \multicolumn{2}{c}{(mas)}& \multicolumn{2}{c}{(mas/yr)}  & \multicolumn{2}{c|}{(mas/yr)}   & \multicolumn{2}{c}{(km/s)}&  &  &Reference for $V_r$\\
\hline
\noalign{\smallskip}
15341& 8.50&1.33& 22.79&1.43&-43.58&1.30&9.21&0.86&21.06&0.97&-43.68&0.70&-1.000&3.600& & &Gontcharov (2006)\\
16407& 7.62&1.15& 22.07&1.30&-46.56&1.03&6.75&0.85&23.22&1.06&-46.91&0.82&-2.347&4.016&+&+&Kunder et al. (2017)\\
16423& 8.44&1.45& 24.13&1.80&-49.19&1.50&8.20&1.32&24.58&1.56&-47.50&1.44&2.150&2.460&+&+&Kunder et al. (2017)\\
16635$^*$& 9.62&2.18& 23.02&2.81&-43.05&2.19&7.95&2.15&20.65&2.84&-49.58&2.45&3.800&0.400& & &Gontcharov (2006)\\
16639$^*$& 8.11&1.47& 19.64&1.80&-43.71&1.31&6.58&1.38&20.58&1.91&-44.51&1.48&5.923&1.952& & &Kunder et al. (2017)\\
16753$^*$& 9.98&1.58& 23.92&1.76&-44.12&1.54&8.17&1.29&22.87&1.63&-44.60&1.65&5.600&0.400& & &Gontcharov (2006)\\
16979$^*$& 5.86&1.77& 20.33&1.87&-43.05&1.69&6.08&1.82&21.34&2.24&-42.16&2.17&6.800&0.780& & &Mermilliod et al. (2009)\\
17000& 7.88&1.00& 19.34&0.98&-45.29&0.90&8.12&0.51&19.88&0.63&-45.25&0.57&4.700&2.200& & &Gontcharov (2006)\\
17034& 6.87&1.08& 21.36&1.24&-46.01&0.85&8.32&0.79&23.91&0.97&-45.11&0.74&3.000&3.000&+& &Kharchenko et al. (2007)\\
17043& 7.78&0.98& 21.89&1.13&-41.60&1.10&7.33&0.61&20.57&0.81&-42.74&0.86&5.900&7.400& & &Kharchenko et al. (2007)\\
17091$^*$& 9.97&1.82& 23.15&2.43&-46.14&1.64&11.82&1.94&26.82&2.74&-44.23&2.11&3.400&0.400& & &Gontcharov (2006)\\
17125& 7.69&1.51& 19.70&1.76&-44.38&1.15&9.19&1.66&21.31&1.90&-46.36&1.67&0.014&2.835& & &Kordopatis et al. (2013)\\
17225& 9.21&1.45& 22.49&1.75&-44.32&1.46&8.10&1.06&21.78&1.42&-44.97&1.26&3.000&7.400& & &Kharchenko et al. (2007)\\
17245$^*$& 5.91&1.67& 14.95&1.73&-46.42&1.50&6.64&1.51&14.67&1.80&-47.59&1.67&3.600&1.300& & &Gontcharov (2006)\\
17289$^*$& 7.29&1.50& 19.70&1.85&-41.66&1.45&7.65&1.50&20.04&1.89&-42.56&1.51&5.721&1.417& & &Kunder et al. (2017)\\
17316$^*$& 6.28&1.66& 24.13&1.70&-47.88&1.22&7.27&1.59&23.97&1.78&-46.89&1.40&7.400&0.600& & &Gontcharov (2006)\\
17401& 9.48&1.11& 20.36&1.15&-45.10&0.97&7.58&0.90&18.77&1.06&-46.36&0.95&-46.900&1.900& &+&Gontcharov (2006)\\
17489& 9.75&1.05& 20.73&0.96&-44.00&0.74&8.65&0.36&20.38&0.43&-44.81&0.37&5.500&0.900& &+&Gontcharov (2006)\\
17497$^*$& 9.76&1.29& 22.71&1.41&-43.67&0.95&8.33&1.22&21.87&1.37&-43.18&1.08&6.230&0.630& & &Mermilliod et al. (2009)\\
17499& 8.80&0.89& 21.55&0.85&-44.92&0.64&8.06&0.25&20.84&0.28&-46.06&0.23&6.700&1.400& &+&Gontcharov (2006)\\
17511$^*$&10.00&1.64& 18.31&1.71&-42.76&1.18&10.67&1.37&16.52&1.46&-43.53&1.15&5.800&0.400& & &Gontcharov (2006)\\
17527& 8.87&0.89& 19.03&0.85&-46.64&0.75&7.97&0.37&20.36&0.45&-46.52&0.41&4.800&0.800& &+&Gontcharov (2006)\\
17531& 8.75&1.08& 19.35&0.95&-41.63&0.76&7.97&0.33&21.24&0.38&-40.56&0.35&7.800&0.600&+&+&Gontcharov (2006)\\
17547& 8.27&1.14& 21.08&1.86&-49.00&1.03&8.82&0.79&20.83&1.56&-48.34&0.92&-0.900&2.200& & &Gontcharov (2006)\\
17552&11.21&1.09& 19.02&1.02&-47.65&0.78&11.04&0.93&18.41&1.00&-46.82&0.87&5.900&2.900& & &Gontcharov (2006)\\
17579& 8.43&0.89& 19.44&0.86&-45.36&0.67&8.77&0.54&20.18&0.70&-44.87&0.62&6.000&0.600& &+&Gontcharov (2006)\\
17583$^*$& 8.50&1.17& 18.88&1.18&-46.40&1.13&8.00&0.89&19.00&0.99&-47.23&0.94&6.500&2.000& & &Gontcharov (2006)\\
17588& 9.21&0.92& 19.83&0.92&-44.38&0.67&8.58&0.56&19.88&0.73&-44.37&0.65&6.900&1.300& &+&Gontcharov (2006)\\
17608& 9.08&1.04& 21.17&0.87&-42.67&0.61&8.58&0.37&21.13&0.35&-43.65&0.27&6.200&2.000&+&+&Wilson (1953)\\
17625& 4.73&1.48& 20.07&1.50&-46.07&1.16&4.42&1.48&20.91&1.65&-45.13&1.38&4.000&4.600& & &Gontcharov (2006)\\
17664& 6.66&0.99& 21.47&0.93&-45.43&0.70&7.66&0.66&22.73&0.84&-45.00&0.85&9.100&0.800&+&+&Gontcharov (2006)\\
17692$^*$& 8.35&1.00& 19.78&0.97&-43.97&0.65&8.90&0.77&18.56&0.75&-44.31&0.60&4.900&0.800& & &Gontcharov (2006)\\
17702& 8.87&0.99& 19.35&0.82&-43.11&0.59&8.09&0.42&19.34&0.39&-43.67&0.33&5.400&1.200&+&+&Gontcharov (2006)\\
17704& 9.05&0.97& 17.84&0.96&-44.94&0.67&9.42&0.75&18.33&0.87&-44.69&0.74&5.000&3.000& & &Gontcharov (2006)\\
17729& 7.61&1.17& 19.34&1.05&-46.91&0.79&9.68&0.93&19.26&0.96&-46.75&0.91&5.100&2.500& & &Gontcharov (2006)\\
17776$^*$& 9.64&0.91& 19.14&0.84&-46.80&0.59&8.45&0.39&17.99&0.39&-46.57&0.32&7.600&0.500& & &Gontcharov (2006)\\
17851& 8.42&0.86& 18.71&0.76&-46.74&0.58&8.54&0.31&18.07&0.30&-47.20&0.27&5.100&0.200&+& &Gontcharov (2006)\\
17862& 8.02&0.91& 18.34&0.86&-44.53&0.67&8.18&0.59&17.42&0.65&-45.38&0.52&7.200&0.900&+&+&Gontcharov (2006)\\
17892&10.12&1.04& 17.80&1.04&-45.00&0.79&8.30&0.66&18.52&0.80&-42.87&0.65&3.800&3.000& & &Gontcharov (2006)\\
17900$^*$& 8.58&0.93& 16.50&0.86&-44.53&0.65&8.72&0.60&16.73&0.63&-44.82&0.53&9.500&0.400& & &Gontcharov (2006)\\
17921&10.14&0.90& 23.88&0.88&-45.90&0.69&8.86&0.42&24.31&0.48&-44.46&0.39&4.100&3.200& & &Gontcharov (2006)\\
17999$^*$& 9.83&1.00& 16.80&0.94&-45.76&0.76&9.93&0.75&18.86&0.83&-43.51&0.69&4.500&0.900& & &Gontcharov (2006)\\
18050$^*$& 7.56&1.47& 20.19&1.42&-45.36&1.04&7.65&1.34&21.84&1.40&-45.50&1.08&9.200&1.800& & &Gontcharov (2006)\\
18091$^*$& 7.71&1.89& 13.35&2.65&-46.20&2.53&6.16&1.42&14.35&2.07&-45.32&1.95&7.600&0.400& & &Gontcharov (2006)\\
18154& 8.57&1.57& 15.43&1.74&-45.83&1.30&10.13&1.66&15.64&1.99&-46.22&1.66&8.000&7.400& & &Kharchenko et al. (2007)\\
18431& 8.66&1.53& 16.99&1.28&-47.06&1.17&7.18&1.48&16.59&1.22&-47.78&1.08&13.000&5.100& & &Kharchenko et al. (2007)\\
18955$^*$& 6.13&1.42& 19.18&1.70&-45.37&1.28&5.88&1.26&20.08&1.54&-46.67&1.12&5.040&1.650& & &Mermilliod et al. (2009)\\
19171$^*$& 6.60&0.85& 22.13&0.91&-50.18&0.79&7.76&0.36&21.88&0.39&-52.34&0.33&-2.000&2.000& & &Wilson (1953)\\

\hline
\label{tab1}
\end{tabular}
\end{table*}

\clearpage
\begin{table*}
\footnotesize
\tabcolsep=0.07cm
\caption{Astrometric parameters for 35 TGAS+RAVE DR5 Pleiades cluster stars.
The stars whose TGAS source IDs are marked with asterisks are used for calculating the apex.
The values of $V_r$ and $\sigma_{V_r}$ are taken from RAVE5 (Kunder et al. 2017). 
}
\vspace{0.5cm}
\centering
\begin{tabular}{lccccccccrccc}
\hline\hline
\noalign{\smallskip}
\multicolumn{1}{c}{Source ID}& RA (J2015.0) & DEC (J2015.0)    & $\pi$ &$\sigma_\pi$ & $\mu_{\alpha}cos(\delta)$ & $\sigma_{\mu_{\alpha}}$ & $\mu_{\delta}$ & $\sigma_{\mu_{\delta}}$ & $V_r$ & $\sigma_{V_r}$ & d1 & d2 \\
\multicolumn{1}{c}{TGAS}  & h m s      & $^\circ$ $'$ $''$& (mas) & (mas)       &  (mas/yr)                 & (mas/yr)                & (mas/yr)       & (mas/yr)                & (km/s)&   (km/s)       &    &    \\
\hline
\noalign{\smallskip}
69335615566555904&052.8165708441&+25.2552705893&7.57&0.34&22.687&0.058&-46.666&0.029&-2.347&4.016&+&+\\
61519668439604992&052.8682026258&+21.8217171596&7.81&0.26&24.116&0.102&-47.109&0.049&2.150&2.460& &+\\
61554646652580736&053.2743659435&+22.1340601259&7.44&0.41&22.813&1.105&-46.278&0.562&0.593&2.512& & \\
69250231614573056$^*$&053.5076283919&+24.8807182731&7.43&0.28&21.268&0.167&-44.702&0.098&3.389&1.319& & \\
68097015715726208$^*$&053.5305604957&+24.3442575792&7.22&0.24&21.637&0.135&-45.051&0.066&5.923&1.952& & \\
67618281484716544&053.8821550266&+22.8233960822&7.67&0.24&21.481&0.148&-45.391&0.084&109.263&2.243& &\\ 
68444873706967808$^*$&054.7370294248&+24.5696285178&7.26&0.23&20.140&0.282&-43.468&0.156&4.952&1.743& & \\
65113559634339200$^*$&054.9216587828&+23.2906992855&7.23&0.23&21.244&0.196&-43.681&0.097&3.921&1.802& & \\
71371258264471424&055.0129405895&+27.7403288914&7.13&0.26&21.011&0.142&-44.371&0.093&0.014&2.835& & \\
70941383577307392&055.0240549596&+26.1961664291&7.87&0.24&21.917&0.740&-49.352&0.449&-8.191&2.534& & \\
68334235349446528&055.1281041510&+24.4871260673&7.27&0.48&23.272&1.463&-44.394&0.752&3.879&2.355& & \\
70190245337962368$^*$&055.1495259881&+26.1512253790&6.64&0.25&18.568&0.815&-40.566&0.338&1.452&1.703& & \\
65150943028579200$^*$&055.3660378253&+23.7081498703&7.72&0.24&20.093&0.753&-45.571&0.371&4.093&1.578& & \\
70108469159560448$^*$&055.4007682128&+25.6191446404&7.66&0.29&16.882&0.107&-43.604&0.064&2.210&1.373& & \\
65027591568709632$^*$&055.5197772040&+22.8583706917&7.36&0.25&21.049&0.138&-44.653&0.075&5.721&1.417& & \\
64317994252099840$^*$&055.6000729045&+21.4732909108&7.84&0.23&20.991&0.150&-47.972&0.094&6.298&1.060& & \\
64449729487990912$^*$&055.6001911465&+22.4209538241&7.33&0.24&19.649&0.195&-44.357&0.094&4.791&1.855& & \\
64879398017459072$^*$&056.2135743548&+23.2687368527&7.68&0.29&20.520&0.092&-44.280&0.060&5.639&1.448& & \\
64739244643463552$^*$&056.2455949969&+22.0322699002&6.98&0.29&18.675&0.111&-43.663&0.061&6.491&1.403& & \\
65275497080596480&056.2773322009&+24.2633219432&8.14&0.46&19.282&0.508&-42.438&0.559&6.760&2.852& & \\
70242781377368704$^*$&056.2842877936&+26.2923245282&7.58&0.48&19.650&0.173&-45.539&0.103&1.169&1.855& & \\
69872039800655744&056.4965089223&+25.3984028056&7.57&0.34&20.288&0.054&-45.367&0.035&11.350&4.254& & \\
69964879813013248&056.8370864639&+25.5256515515&7.63&0.30&18.481&0.797&-44.980&0.442&2.653&2.232& & \\
64898364591843712&056.8455075529&+22.9219212894&6.62&0.66&21.620&0.054&-47.038&0.034&6.791&2.549& &+\\
69948249699646720&056.9452369007&+25.3855062674&7.37&0.28&19.469&0.070&-45.724&0.049&-1.840&3.861& & \\
64933755122821120&057.1829845987&+23.2596275538&7.01&0.30&19.376&0.431&-44.386&0.303&4.465&2.279& & \\
66980358578521856$^*$&057.4705683282&+25.6472592675&7.17&0.24&18.072&1.124&-45.697&0.609&2.535&1.833& & \\
64924409273987712$^*$&057.4854860471&+23.2184434833&6.61&0.40&20.067&1.469&-42.796&0.373&4.336&1.495& & \\
66558249192653952&057.4859034805&+24.3488065455&7.41&0.35&19.031&0.322&-44.427&0.330&8.457&9.470& & \\
66960258131598720$^*$&057.5737256782&+25.3793765749&7.35&0.25&19.397&1.011&-45.891&0.501&2.957&1.201& & \\
64172034082472448&057.5888577436&+23.0961577837&7.20&0.61&20.869&2.236&-46.430&0.505&4.174&2.344& & \\
63730305286697600$^*$&057.9254636097&+21.6681735475&7.22&0.27&18.369&1.411&-43.122&0.657&3.654&1.329& & \\
66570549979009280&058.3488657043&+24.0648492826&7.33&0.30&19.792&1.035&-45.735&0.595&10.642&3.779& & \\
65819961494790400&058.5900341254&+24.0755010734&7.71&0.26&19.458&0.872&-45.686&0.447&-3.823&5.458& & \\
65754128236100096&059.1092500687&+23.7841281330&6.64&0.24&22.523&1.351&-41.646&0.550&44.166&3.392& & \\
\hline
\label{tab2}
\end{tabular}
\end{table*}

\clearpage
\begin{table*}
\small
\tabcolsep=0.1cm
\caption{Our kinematical parameters corresponding to HIP New (19 stars) ) 
and TGAS+RAVE DR5 (17 stars from Table~2) data with other published results.}
\vspace{0.5cm}
\centering
\begin{tabular}{lll}
\hline\hline
Parameter              &             Results             &        Reference \\
\hline
\noalign{\smallskip}
$\overline{V_x}, \overline{V_y},\overline{V_z}$  km/s  & $-$1.07, 20.26, $-$23.26   & Table~1    \\
$\overline{V_x}, \overline{V_y},\overline{V_z}$  km/s  & $-$1.94, 19.98, $-$24.70   & Table~2  \\
$A_{CP}, D_{CP}$, degree               & $95.73 \pm 3.56$, $-50.44 \pm 8.84$ & Table~1 \\
$A_{CP}, D_{CP}$, degree               & $92.52 \pm 1.72$, $-42.28 \pm 2.56$ & Table~2 \\
$A_0, D_0$, degree                            & $93.06 \pm 5.95$, $-48.42 \pm 4.02$ & Table~1 \\
$A_0, D_0$, degree                            & $95.59 \pm 2.30$, $-50.90 \pm 2.04$ & Table~2 \\
$A_0, D_0$, degree                            &  89.7, $-$35.15                & Montes et al. (2001) \\
$A_{CP}, D_{CP}$, degree               &  $92.49 \pm 5.4$, $-47.87 \pm 5.3$   & Makarov \& Robichon (2001)  \\
$A_0, D_0$, degree                            &  $92.9 \pm 1.2$, $-49.4 \pm 1.2$   & Galli et al. (2017)     \\
$V$, km/s                                           &  35.36                             & Table~1 \\
$V$, km/s                                           &  26.72                             & Table~2 \\
$V$, km/s                                           &  26.5                               & Montes et al. (2001) \\
$d$, pc                                                &  $124.5 \pm 24.1$          & Table~1 \\
$d$, pc                                                &  $136.8 \pm 6.4$            & Table~2 \\
$d$, pc                                                &  $134 \pm 6$                  & Gaia Collaboration et al. (2016) \\
$d$, pc                                                &  $132.0 \pm 2.0$            & Munari et al. (2004)    \\
$d$, pc                                                &  $134.6 \pm 3.1$            & Soderblom et al. (2005) \\
$d$, pc                                                &  $122.2 \pm 1.9$            & van Leeuwem (2007)      \\
$d$, pc                                                &  $134.4^{+2.9}_{-2.8}$   &  Galli et al. (2017)    \\
$d$, pc                                                &  $134.8\pm1.7$                &  M\"adler et al. (2016)   \\
$(m-M)$                                              &  $5.48 \pm 0.32$             & Table~1 \\
$(m-M)$                                              &  $5.68 \pm 0.12$             & Table~2 \\
$(m-M)$                                              &  $5.60$                           & Munari et al. (2004)    \\
$(m-M)$                                              &  $5.65 \pm 0.05$            & Soderblom et al. (2005) \\
$(m-M)$                                              &  $5.44 \pm 0.03$            & van Leeuwem (2007)      \\
$\overline{U}, \overline{V}, \overline{W}$, km/s       & $-6.38\pm0.32$, $-26.91\pm2.04$, $-13.69\pm0.16$ & Table~1 \\
$\overline{U}, \overline{V}, \overline{W}$, km/s       & $-5.39\pm0.12,$ $-28.29\pm1.36$, $-13.53\pm1.16$ & Table~2 \\
$\overline{U}, \overline{V}, \overline{W}$, km/s       & $-11.6, -21.0, -11.4$                & Montes et al. (2001)    \\
$\overline{U}, \overline{V}, \overline{W}$, km/s       & $-6.2, -28.7, -14.7$                 & Galli et al. (2017)     \\
space velocity $V=\sqrt{\overline{U}^2+\overline{V}^2+\overline{W}^2}$ &  $30.86 \pm 3.72$  & Table~1 \\
space velocity $V=\sqrt{\overline{U}^2+\overline{V}^2+\overline{W}^2}$ &  $32.14 \pm 2.24$ & Table~2 \\
space velocity $V=\sqrt{\overline{U}^2+\overline{V}^2+\overline{W}^2}$ &  $32.9 \pm 0.3$    & Galli et al. (2017)  \\
$x_c, y_c, z_c$, pc                                    &  64.97, 98.30, 51.37           & Table~1   \\
$x_c, y_c, z_c$, pc                                    &  70.36, 103.58, 55.60           & Table~2  \\
$\lambda_1, \lambda_2, \lambda_3$, km/s                &  931.92, 6.30, 2.94             & Table~1  \\
$\lambda_1, \lambda_2, \lambda_3$, km/s                &  954.54, 1.47, 0.35             & Table~2  \\
$\sigma_1, \sigma_2, \sigma_3$, km/s                   &  30.53, 2.51, 1.72             & Table~1   \\
$\sigma_1, \sigma_2, \sigma_3$, km/s                   &  30.90, 1.21, 0.59              & Table~2 \\
$l_1, m_1, n_1$, degree                                &  0.20, 0.87, 0.44               & Table~1   \\
$l_1, m_1, n_1$, degree                                &  0.17, 0.89, 0.42               & Table~2 \\
$l_2, m_2, n_2$, degree                                & $-0.87, 0.36, -0.33$             & Table~1  \\
$l_2, m_2, n_2$, degree                                & $-0.91, 0.30, -0.25$             & Table~2  \\
$l_3, m_3, n_3$, degree                                &  0.45, 0.32, $-0.84$             & Table~1   \\
$l_3, m_3, n_3$, degree                                &  0.36, 0.35, $-0.87$             & Table~2 \\
\hline
\label{tab3}
\end{tabular}
\end{table*}

\clearpage
\begin{table*}
\small
\tabcolsep=0.1cm
\caption{Description of the number of stars in different panels of Fig.~1 (black and red points).}
\vspace{0.5cm}
\centering
\begin{tabular}{lcccc}
\hline\hline
panel&  \multicolumn{2}{c}{left, single stars} &  \multicolumn{2}{c}{right, double and multiple stars}  \\
\cline{2-5}
signature,&$\sigma_{V_r}\le2.0$ km/s,&$\sigma_{V_r}>2.0$ km/s,&$\sigma_{V_r}\le2.0$ km/s,&$\sigma_{V_r}>2.0$ km/s,\\
data&red& black&red& black\\
\hline
(a), HIP New& 19 & 14 & 12 & 3 \\
(b), HIP       &  19 & 14 & 12 & 3 \\
(c), TGAS   &  17  & 13 &  0 & 3 \\
\hline
\label{tab4}
\end{tabular}
\end{table*}


\clearpage
\begin{figure*}[]
\centering
\vspace{-0.7cm}
\includegraphics[width=14cm]{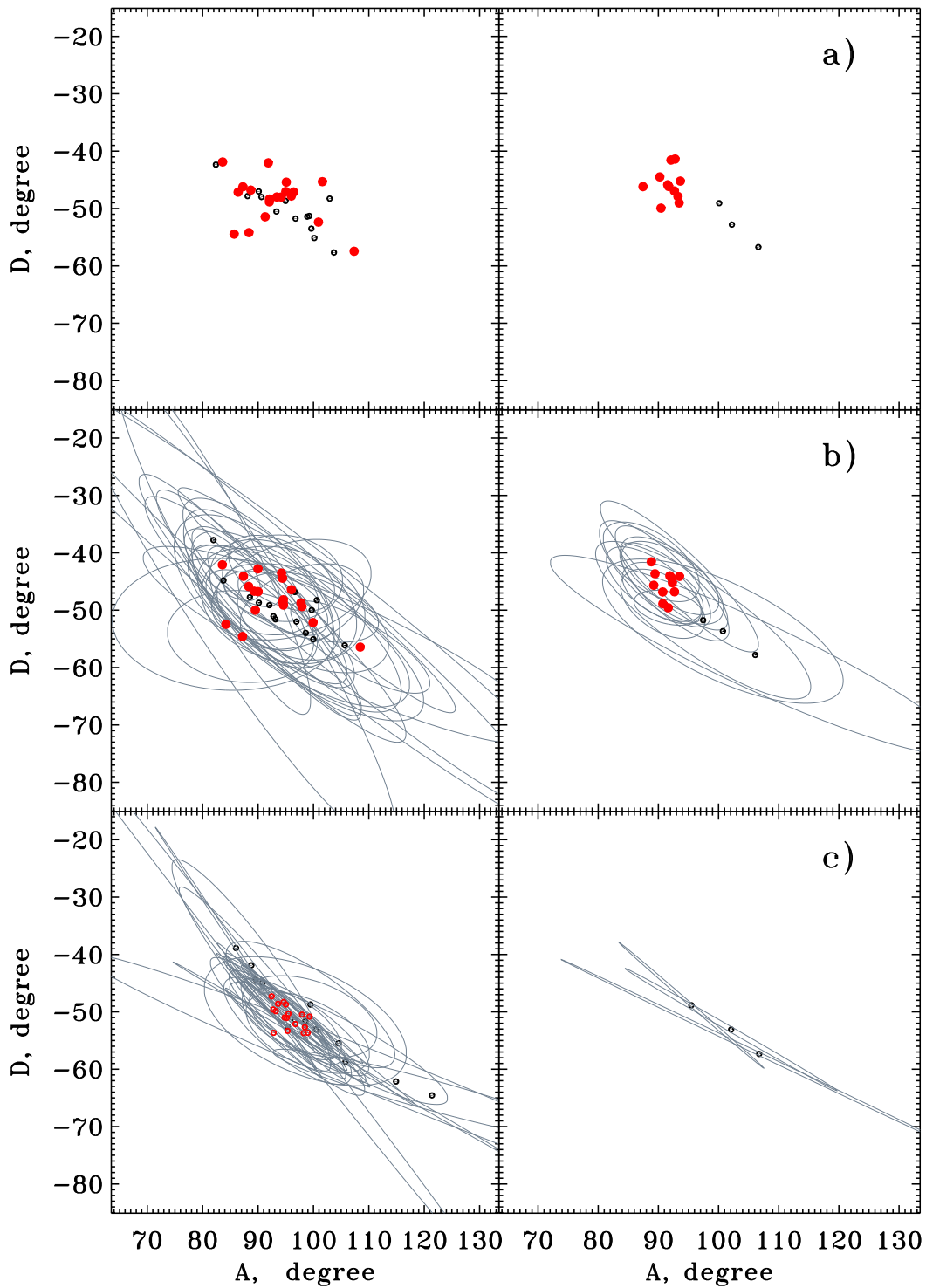}
\caption{ 
The AD-diagrams based on various data: 
(a) HIP New, (b) HIP and (c) TGAS+RAVE DR5.  
The left panels for all three data sets are for single stars
and right panels are for double and multiple stars.
The red dots denote stars whose radial velocity errors do not exceed 2~km/s. 
Error ellipses are plotted for both panel (b) and (c), i.e. for HIP and TGAS data.
The left panel (c), for TGAS stars, does not show the star 67618281484716544 with $V_r = 109.263$~km/s  
and 65754128236100096 with $V_r = 44.166$~km/s, 
whose positions are outside the limits of the figure. 
Also, in the left panel of (c), the number of single stars with  $\sigma_{V_r}>$2~km/s
is 13, instead of 15. 
}
\label{fig1}
\end{figure*}

\clearpage
\begin{figure*}[]
\centering
\vspace{-0.7cm}
\includegraphics[width=10cm,height=10cm]{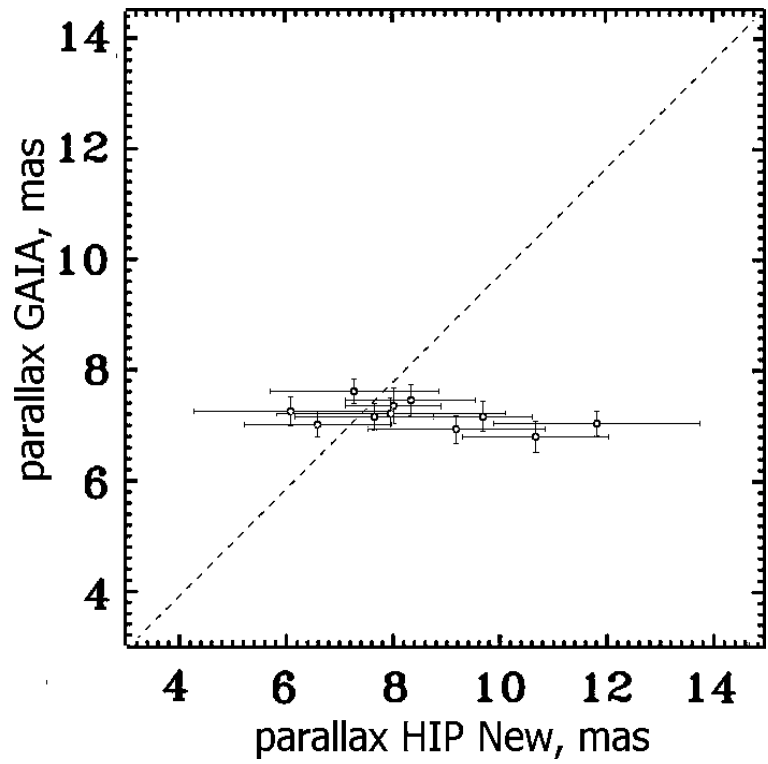}
\caption{Comparison of TGAS parallaxes for HIP New stars (single stars from Table 1,  
with HIP IDs 16635, 16639, 16979, 17091, 17125, 17289, 17316, 17497, 17511, 17583 and 17729). 
}
\label{fig2}
\end{figure*}

\end{document}